\documentclass[prl,amsmath,amssymb,natbib]{revtex4}
\usepackage{amssymb,multirow,color,graphicx}

\renewcommand{\vec}[1]{\bm{#1}}
\renewcommand{\vec}[1]{\mathbf{#1}}
\begin{document}

\title{The statistics of a passive scalar in field-guided magnetohydrodynamic turbulence}
\author{J. Mason$^1$, S. Boldyrev$^2$, F. Cattaneo$^3$, J.C. Perez$^4$}
\affiliation{${~}^1$College of Engineering, Mathematics and Physical Sciences, University of Exeter, EX4 4QF, UK\\
${~}^2$Department of Physics, University of Wisconsin at Madison, 1150 University Ave, Madison, WI 53706, USA\\
${~}^3$Department of Astronomy \& Astrophysics, University of Chicago, 5640 S. Ellis Ave, Chicago, IL, 60637, USA\\
${~}^4$Institute for the Study of Earth, Oceans, and Space, University of New Hampshire, Morse Hall, 8 College Road, Durham, NH, 03824\\
{\sf j.mason@exeter.ac.uk},  {\sf boldyrev@wisc.edu}, {\sf cattaneo@flash.uchicago.edu}, {\sf jeanc.perez@unh.edu}}

\begin{abstract}
A variety of studies of magnetised plasma turbulence invoke theories for the advection of a passive scalar by turbulent fluctuations. Examples include modelling the electron density fluctuations in the interstellar medium, understanding the chemical composition of galaxy clusters and the intergalactic medium, and testing the prevailing phenomenological theories of magnetohydrodynamic turbulence. While passive scalar turbulence has been extensively studied in the hydrodynamic case, its counterpart in MHD turbulence is significantly less well understood. Herein we conduct a series of high-resolution direct numerical simulations of incompressible, field-guided, MHD turbulence in order to establish the fundamental properties of passive scalar evolution. We study the scalar anisotropy, establish the scaling relation analogous to Yaglom's law, and measure the intermittency of the passive scalar statistics. We also assess to what extent the pseudo Alfv\'en fluctuations in strong MHD turbulence can be modelled as a passive scalar.  The results suggest that the dynamics of a passive scalar in MHD turbulence is considerably more complicated than in the hydrodynamic case.\\
\end{abstract}


\maketitle

\section{Introduction}
The evolution of a passive scalar field in hydrodynamic turbulence is a classical problem that is widely studied. The scalar typically represents a chemical contaminant or admixture that has no dynamical effect on the flow, or small temperature fluctuations in which buoyancy effects are negligible. Indeed, scalar turbulence is important for understanding the dispersal of a pollutant in atmospheric flows, mixing problems in ocean dynamics, and reaction and combustion problems in chemical engineering \citep{warhaft_00,shraiman_s00,aref_etal14, dimotakis_05,jacobson_01,pitsch_06}. A concentration of solute also provides a means of flow visualisation in experimental fluid dynamics. In the years since the pioneering studies of \cite{kolmogorov_41a}, \cite{obukhov_49},  \cite{corrsin_51}, \cite{yaglom_49} and \cite{batchelor_59} a tremendous range of theoretical, computational and experimental investigations have addressed the scalar statistics at the integral, inertial and dissipative scales \citep[see the reviews by][and references therein]{warhaft_00,shraiman_s00,dimotakis_05,aref_etal14}. 

Turbulent mixing also plays an important role in astrophysical systems.  The chemical composition of the intergalactic medium and of galaxy clusters is affected by the turbulent nature of the plasma \citep{friel_b92,schaye_etal03,pichon_etal03,rebusco_etal05,pieri_etal06,desilva_etal06,bruggen_s09}. The interpretation of scintillation measurements of distant radio sources crucially depends on fluctuations of the electron density that is possibly mixed by interstellar turbulence \citep{rickett90,gwinn_etal93,lithwick_g01,boldyrev_g03}. Density fluctuations in the solar wind at scales much larger than the plasma microscales (such as the ion gyroscale and ion inertial length) are consistent with being passively advected from large to small scales by magnetohydrodynamic turbulence. 

An important property of astrophysical systems is the presence of magnetic fields that can be generated by the turbulent motion of the highly conductive plasma, through a process known as dynamo action  \cite[e.g.,][]{tobias_cb11,brandenburg_etal12}. One of the most important outstanding problems of MHD turbulence is understanding how ordered large-scale magnetic fields are generated out of the small-scale state. Herein, rather than focussing on the dynamo problem, we concentrate on passive scalar MHD turbulence in the presence of a strong uniform background magnetic field. There are a number of reasons for this choice. First, considerable progress has been made recently with the theory of strong field-guided MHD turbulence \citep{goldreich_s95, maron_g01, muller_bg03, boldyrev_06, perez_etal12}. These advances have also highlighted a number of issues related to passive scalar evolution, including a description of the dynamically less significant pseudo Alfv\'en fluctuations as being passively advected by the shear Alfv\'en fields. The presence of residual energy in MHD turbulence and the associated different scalings for the kinetic, magnetic and total energy of the system \citep{boldyrev_pbp11, chen_etal13} also raises the question of whether the spectrum of a passive scalar will follow that of the advecting flow. Moreover, justification for studying the field-guided regime is found by noting that even in the absence of a large-scale magnetic field, locally the magnetic field that is created by the large scale turbulent eddies will play the role of a guiding magnetic field for the small-scale plasma fluctuations. We note that a numerical study of passive scalar advection in a setting without a large-scale magnetic field has been recently performed by \cite{sur_etal2014}. The spectrum of a passive scalar in the field-guided case has also been previously addressed in the lower resolution simulations of \cite{maron_g01}, and the related issue of the statistics of tracer particles in field-guided MHD turbulence has been studied by \cite{busse_m08}. Our present work complements these studies.     

We conduct a series of high-resolution direct numerical simulations of strong, forced, incompressible, field-guided MHD turbulence in which an additional driven passive scalar field is evolved. We establish that, since the turbulence is critically balanced, the passive scalar satisfies a Yaglom relation which takes into account the anisotropy with respect to the background magnetic field. Through studying the properties of the scalar in the statistically steady regime, we analyse the scalar spectrum, anisotropy and intermittency. We also investigate whether the dynamically less significant pseudo Alfv\'en fields are adequately described by a passive scalar field, and we address the extent to which their statistics are similar to those of the dominant shear Alfv\'en fluctuations.

\section{Numerical simulations}
The equations describing the evolution of a passive scalar field in incompressible MHD turbulence read 
\begin{subequations}
\label{eq:mhd}
\begin{equation}
\label{eq:mhd_u}
\frac{\partial \vec{u}}{\partial t} +\left(\vec u \cdot \nabla \right)\vec u = -{\nabla} p + (\nabla \times \vec{B}) \times \vec{B} +\nu\nabla^2 \vec u+\vec f^u_\perp, 
\end{equation}
\begin{equation}
\frac{\partial \vec{B}}{\partial t}   = \nabla \times (\vec{u} \times \vec{B}) +\eta\nabla^2 \vec B +\vec f^B_\perp,  
\end{equation}
\begin{equation}
\label{eq:mhd_div}
\nabla \cdot {\vec u}=0, \quad \nabla \cdot {\vec B}=0, 
\end{equation}
 \end{subequations}
\begin{equation}
  \label{eq:scalar}
\frac{\partial s}{\partial t}+(\vec{u} \cdot \nabla ) s = \kappa \nabla^2 s +f^s,
  \end{equation}
where $\vec{u(x},t)$ is the velocity, $\vec{B}(\vec{x},t)$ is the magnetic field (measured in units of the Alfv\'en speed $B/\sqrt{4\pi\rho_0}$), $s(\vec{x},t)$ is the passive scalar quantity, $p$ is the pressure, and $\nu$, $\eta$ and $\kappa$ are the fluid viscosity, the magnetic diffusivity and the scalar diffusivity, respectively. Henceforth, the magnetic field will be decomposed into the uniform background magnetic field and the fluctuations,  $\vec{B}(\vec{x},t)=B_0\vec{\hat e_z}+\vec{b}(\vec{x},t)$ where $B_0$ is a constant. The turbulent flow is driven at large scales by the random force $\vec f^u_\perp$. In order to independently drive both Els\"asser populations, $\vec{z^\pm}=\vec{u}\pm\vec{b}$, we have added a force $\vec f^B_\perp$ to the induction equation. In order to measure the properties of the scalar in the statistically steady state we have also added a large scale driving $f^s$ to the equation for the passive scalar.

The design of our numerical calculations is based on our earlier investigations that have established the properties of the simulation setup that are most conducive to studying strong field-guided MHD turbulence. Details of those calculations and a summary of the findings can be found in~\cite{mason_pbc12}. Here, we simply state that  for all of the calculations reported in this paper we have set $B_0=5$ (in units of the rms velocity $u_{rms} \approx 1$) and the domain is elongated in the direction of the background field with the aspect ratio $L_\|/L_\perp=6$ and $L_\perp=2\pi$.  The forces $\vec f^u_\perp$ and $\vec f^B_\perp$ have no component along $z$, are solenoidal in the xy-plane, and are applied in Fourier space at the wavenumbers $1 \le k_{x,y} \, L_\perp/2\pi \le 2 $ and $k_{z} = \pm 2\pi/L_z$. The individual random values are chosen from a Gaussian distribution with a zero mean. The values are refreshed independently on average approximately $10$ times per large-scale eddy turnover time ($L_\perp/u_{rms} \approx 1$) and the ratio of the cross helicity to the total energy $\sigma_c = 2\langle \vec{u} \cdot \vec{b} \rangle/\langle u^2 +b^2\rangle \approx 0$.   The scalar force $f^s$ acts at the same wavenumbers as the Els\"asser driving and has similar statistical properties, except that the scalar amplitudes are updated 10 times more frequently (to ensure that the rate of passive scalar stirring is fixed). 

The equations are solved using standard pseudospectral methods. Unless otherwise stated, the grid resolution is $1024^3$ mesh points, the Reynolds number $Re=1/\nu=5600$, the magnetic Prandtl number $Pm=\nu/\eta=1$ and the P\'eclet number $Pe=1/\kappa=Re$. The numerical results presented correspond to averages over approximately 40 independent snapshots of the system in the statistically steady state.

\subsection{The anisotropic Yaglom relation}

\begin{figure} 
\begin{center}
\resizebox{\textwidth}{!}{\includegraphics{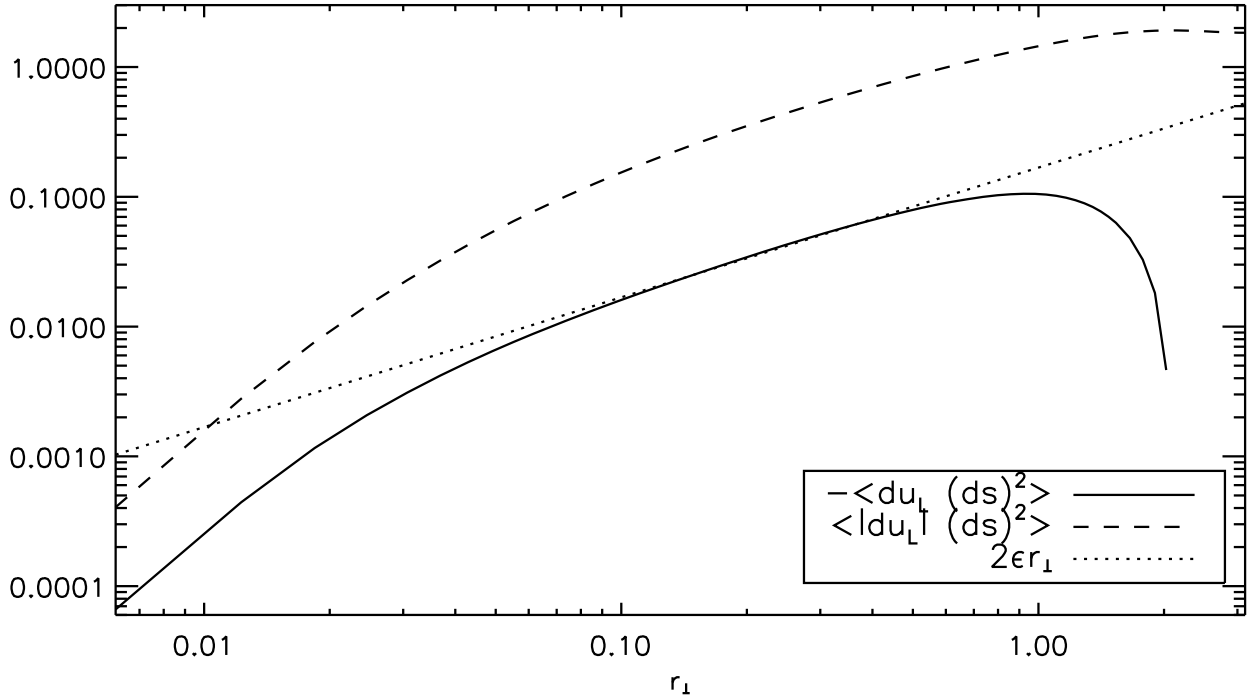}}
\caption{The Yaglom relation (\ref{yaglom2}) as measured from the numerical simulations. Also shown is the same quantity with $\delta  u_L$ replaced by its absolute value. This is sometimes measured in the hydrodynamic turbulence literature as it avoids cancellations and hopefully extends the inertial range, although no significant improvement is observed here.}
\label{fig:yaglom}
\end{center}
\end{figure}

We begin by establishing one of the central relations in turbulent passive scalar advection, the Yaglom relation \cite[][]{yaglom_49}. In the inertial interval of the velocity field, when the passive scalar stirring and diffusion can be neglected, the relation states that
\begin{eqnarray}
\nabla \cdot \langle \delta {\vec u} \, (\delta s)^2 \rangle=-4\epsilon_s,
\label{yaglom1}
\end{eqnarray}
where $\delta  s=s(\vec{x+r})-s(\vec{x})$,  ${\delta {\vec u}}=\vec{u}(\vec{x+r})-\vec{u}(\vec{x})$, and $\epsilon_s$ is the rate of passive scalar dissipation, $\epsilon_s=\kappa \langle (\bm\nabla s)^2 \rangle$. This expression assumes homogeneity of the turbulence but does not require isotropy. In critically balanced anisotropic MHD turbulence we have $\nabla_\| / \nabla_\perp \sim \delta  { u}_\|/\delta \bm{ u}_\perp\ll 1$ \citep{goldreich_s95}, where the subscripts denote the components parallel and perpendicular to the local guiding magnetic field. We can therefore neglect the parallel gradient in expresssion~(\ref{yaglom1}), and if we then take the point separation vector ${\bf r}$ in the field-perpendicular plane where the turbulence is isotropic we obtain
\begin{eqnarray}
\label{yaglom2}
\langle \delta  u_L \, (\delta  s)^2 \rangle=-2\epsilon_s r_\perp.
\end{eqnarray}
This is a generalisation of the isotropic Yaglom relation to anisotropic critically balanced MHD turbulence. Figure~\ref{fig:yaglom} shows the results of computing the left and right-hand sides of expression (\ref{yaglom2}) from our numerical simulations. We note that $\delta  u_L = \delta  \vec{u}{\bm \cdot} \vec r/|\vec r|$ and $\vec r$ is to be taken in the plane perpendicular to the local guiding magnetic field. In the case of a strong background field $\vec B_0=B_0\vec{\hat e_z}$ this is approximately the $xy$-plane. For simplicity we take $\vec r =r_\perp \vec{\hat e_x}$. Figure~\ref{fig:yaglom} illustrates that the simulations reproduce Yaglom's relation quite well and indicate the presence of the inertial interval between $r_\perp \approx 0.07$~and~$0.5$.

\begin{figure}
\begin{center}
\resizebox{\textwidth}{!}{\includegraphics{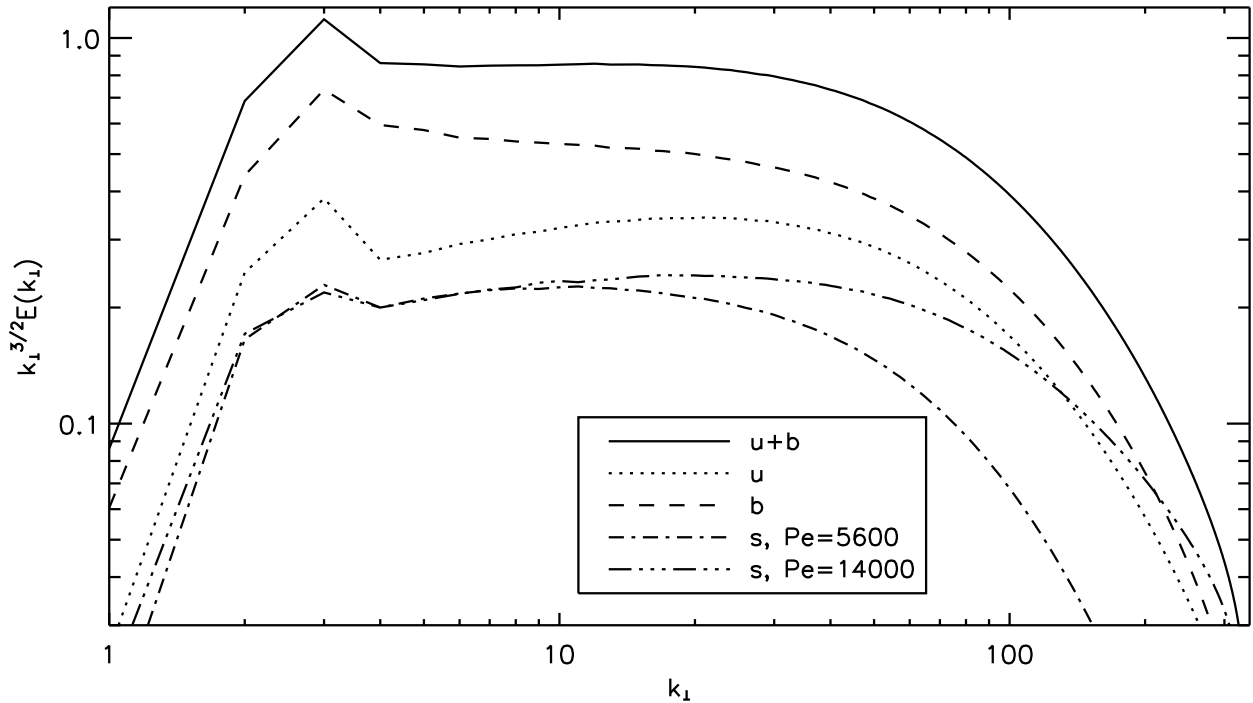}}
\caption{A comparison of the spectrum of the passive scalar with that of the velocity, the magnetic field and the total energy (all of the spectra have been compensated by $k_\perp^{3/2}$). The scalar spectrum is shown for two different values of $Pe$. }
\label{fig:spectra_1024}
\end{center}
\end{figure}

\subsection{The passive scalar spectrum}

In hydrodynamic turbulence, the passive scalar spectrum is expected to follow the spectrum of the energy \citep{batchelor_59, biskamp_03}. However, in MHD turbulence both numerical simulations and physical systems with limited scale separations show that there is a slight but measurable mismatch between the apparent spectra of the kinetic and magnetic energies, the difference being termed the residual energy \citep{pouquet_fl76, boldyrev_pbp11,chen_etal13}. It is therefore of interest to investigate whether the scalar spectrum in MHD turbulence follows that of the flow, the magnetic field or the total energy. We define the field-perpendicular spectrum of the quantity $q$ (where $q$ represents either the velocity, the magnetic field or the scalar) by
\[E_q(k_{\perp})=\tfrac{1}{2} \langle | \vec{\hat q}(k_\perp)|^2 \rangle k_\perp,\]
where $\vec{\hat q}(k_{\perp})$ is the two-dimensional Fourier transformation of
$\vec{q(x)}$ in a plane perpendicular to $\vec{B_0}$ and $k_\perp=\sqrt{k_x^2+k_y^2}$. The average is taken over all field-perpendicular planes in
the data cube and then over all data cubes (i.e., snapshots).

In figure~\ref{fig:spectra_1024} we compare the (compensated) scalar spectrum with that of the kinetic, magnetic and total energy for two different values of the P\'eclet number. The different behaviour of the MHD energies is apparent: the total energy $E_T \propto k_\perp^{-3/2}$, while the velocity has a spectral exponent that is slightly shallower than $-3/2$ and the magnetic spectrum is of higher amplitude and is slightly steeper than $-3/2$.  While it it is difficult to identify the inertial range scaling of the scalar spectrum, especially for the case in which $Pe=Re$, for larger values of the P\'eclet number the scalar spectrum appears to more closely follow that of the velocity. We have observed that the passive scalar requires a P\'eclet number a few times larger than the Reynolds number to exhibit an inertial interval comparable to that of the advecting velocity field.

\subsection{The passive scalar anisotropy}

Since the scalar is advected by a flow in which the field-perpendicular gradients are much larger than the field-parallel ones \citep{goldreich_s95, boldyrev_06}, it is natural to expect that the scalar will be more effectively mixed in the field-perpendicular plane.  In order to characterise the scalar anisotropy we have calculated the typical field-parallel length scale $\l_\|=\sqrt{\langle s^2 \rangle/\langle(\partial s/\partial z)^2\rangle}$ and the corresponding field-perpendicular length scale $\l_\perp=\sqrt{\langle s^2 \rangle/\langle(\partial s/\partial x)^2\rangle}$. Quite remarkably, we find $\l_\perp/\l_\| \approx 1/B_0$, independent of the grid resolution and the anisotropy of the scalar force. Thus the scalar rapidly assumes the anisotropy of the advecting velocity field.\\

\subsection{The pseudo Alfv\'en fluctuations}

We now turn to the issue of the pseudo Alfv\'en fluctuations. We are particularly interested in investigating the similarities and differences with the statistics of the passive scalar field. In strong MHD turbulence the dominant fluctuations have wavenumbers $k_\perp \gg k_\|$, and for large $k_\perp$ the polarisation of the pseudo Alfv\'en fluctuations is almost along the direction of the background field while the shear Alfv\'en waves are polarised perpendicular to $\vec{B_0}$. The equation for the pseudo Alfv\'en fluctuations can therefore be obtained by writing the MHD equations (\ref{eq:mhd_u}-c) in terms of the Els\"asser variables $\vec{z^\pm}=\vec{u} \pm \vec{b}$, and taking the field-parallel component:
\begin{equation}
\label{eq:pA}
\left(\frac{\partial}{\partial t} \mp B_0 \nabla_\| \right) z_\|^\pm+\vec{z_\perp^\mp} \cdot \nabla_\perp z_\|^\pm + z_\|^\mp \nabla_\| z_\|^\pm=-\nabla_\| P+\nu \nabla^2 z_\|^\pm,
\end{equation}
where $P=p/\rho_0+B^2/2$ is the total pressure (for simplicity we have set $\nu=\eta$).  

If we ignore for the moment the right-hand side of equation (\ref{eq:pA}), then the pseudo Alfv\'en cascade is controlled by the two terms $\vec{z_\perp^\mp} \cdot \nabla_\perp z_\|^\pm$ and $z_\|^\mp \nabla_\| z_\|^\pm$. However, the latter can be neglected since $k_\| \ll k_\perp$ 
\citep[from which it also follows that $|z_\|| \lesssim |\vec{z_\perp}|$, see][]{biskamp_03} and we obtain
\begin{equation}
\label{eq:pA_approx}
\left(\frac{\partial}{\partial t} \mp B_0 \nabla_\| \right) z_\|^\pm+\vec{z_\perp^\mp} \cdot \nabla_\perp z_\|^\pm =-\nabla_\| P+\nu \nabla^2 z_\|^\pm \,.
\end{equation}
It is due to the resemblance of this equation with (\ref{eq:scalar}) for the passive scalar (with the velocity replaced by the shear Alfv\'en fluctuations) that it is commonly stated that the spectrum of pseudo Alfv\'en fields must follow that of the `advecting' shear Alfv\'en fields. We note however that the right-hand sides of equations (\ref{eq:scalar}) and (\ref{eq:pA_approx}) differ through the the presence of the forcing term $f_s$ in the scalar equation and the pressure term $-\nabla_\|P$ in the pseudo Alfv\'en equation. Indeed, in order to simulate the dominant shear Alfv\'en cascade of strong field-guided MHD turbulence cost effectively, in numerical simulations it is typically only the shear Alfv\'en components that are forced. Thus, as shown by (\ref{eq:pA_approx}), the only source for the pseudo Alfv\'en fluctuations is the pressure term, which acts at all scales. This is fundamentally different to the source term for the scalar, which acts only at large scales. 

\begin{figure} 
\begin{center}
\resizebox{\textwidth}{!}{\includegraphics{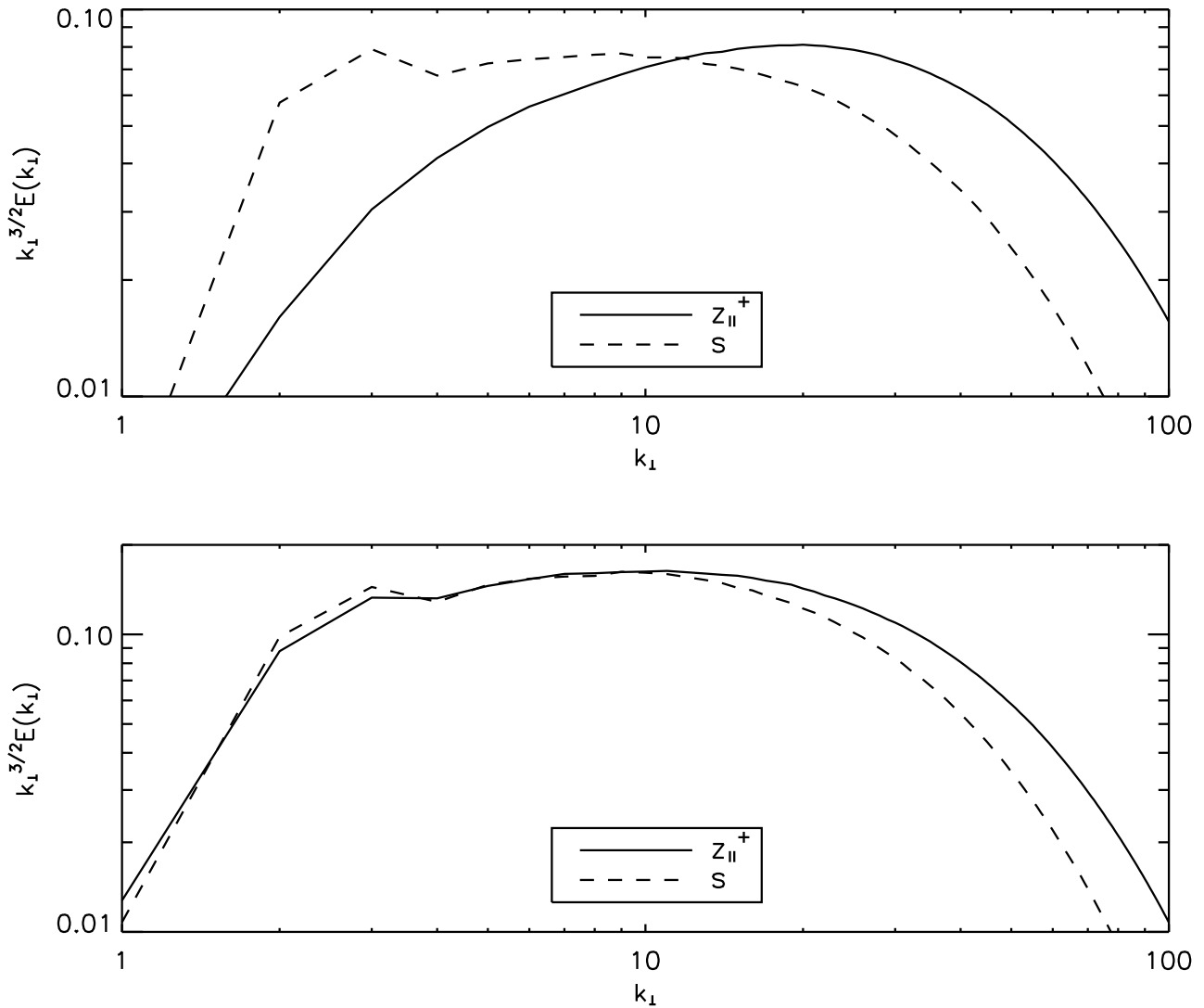}}
\caption{A comparison of the scalar spectrum and the field-parallel Els\"asser spectrum. In the top frame only the field-perpendicular fluctuations are driven. In the bottom frame all three components of the velocity are driven. In both cases the scalar is advected by $\vec u$ ($512^3$, $Re=Pe=1800$).}
\label{fig:pseudo_spec}
\end{center}
\end{figure}

In order to investigate the effects of this difference we have conducted a series of simulations in which we have added field-parallel forces ($f^u_\|$ and $f^B_\|$) to the MHD equations (\ref{eq:mhd}). As shown in the upper panel of figure~\ref{fig:pseudo_spec}, our findings suggest that when the pseudo Alfv\'en field is not externally forced ($f^\pm_\| = f^u_\| \pm f^B_\| =0$) its spectrum is governed by the pressure term, and it is therefore different to the spectrum of a passive scalar. However, when the pseudo Alfv\'en field is forced, the situation is different. The lower panel of figure~\ref{fig:pseudo_spec} illustrates the case when the amplitude of $f^\pm_\|$ is chosen to be a factor of approximately $B_0$ smaller than $|\vec{f^\pm_\perp}|$, so that the total force preferentially excites the shear Alfv\'en waves (the wavenumbers excited by $f^\pm_\|$ and timescale of the force remain the same as those for $\vec{f^\pm_\perp}$). We observe that while the pressure still acts at all scales, as a result of the the field-parallel forcing the pseudo Alfv\'en spectrum more closely resembles that of the scalar. \\

\subsection{The intermittency of the passive scalar}

\begin{figure} 
\begin{center}
\resizebox{\textwidth}{!}{\includegraphics{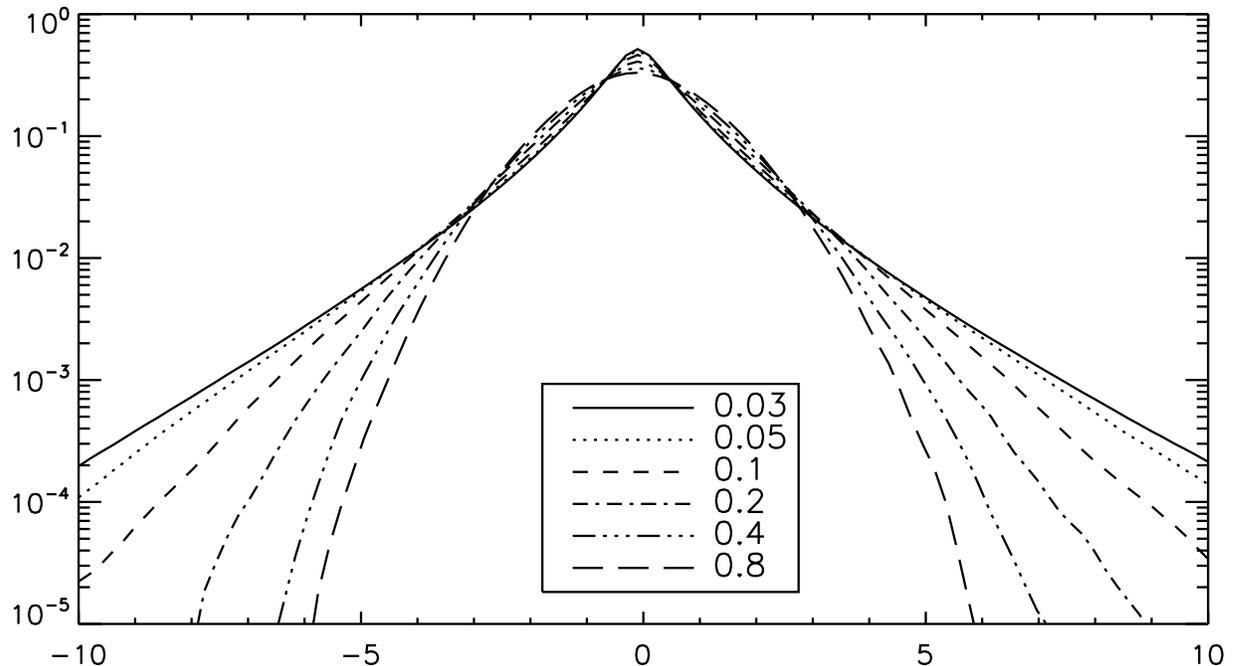}}
\caption{Histogram of the normalised scalar differences $\hat{\delta s}$ for a variety of increments $r$. The histogram for $r=0.8$ is almost indistinguishable from a Gaussian ($512^3$, $Re=Pe=1800$).}
\label{fig:pdf_scalar_r}
\end{center}
\end{figure}

Finally, we investigate the intermittency of the scalar and compare it with that of the shear and pseudo Alfv\'en fluctuations. Shown in figure~\ref{fig:pdf_scalar_r} are the histograms of the normalised scalar differences $\hat{\delta  s}=\delta  s/\langle |\delta  s| \rangle$. The values of the increments $\vec r=r \vec{\hat e_x}$ are intended to be representative of the forcing scales ($r=0.8$), the inertial range ($r=0.4,0.2,0.1$) and the dissipative scales ($r=0.05$). The scalar statistics are Gaussian at large $r$, reflecting the Gaussian forcing mechanism, and the scalar becomes increasingly intermittent at smaller scales. A more qualitative measure of the intermittency is given by the kurtosis $K=F-3$ where $F$ is the flatness $F=\langle{(\hat{\delta s})^4}\rangle/\langle{(\hat{\delta  s})^2}\rangle^2$. For a Gaussian distribution the kurtosis is zero. For the scalar field we obtain the values $K=7.7, 5.8 , 3.4, 1.7, 0.8, 0.3$ for the increments  $r=0.03, 0.05, 0.1, 0.2, 0.4, 0.8$, respectively.

\begin{figure} 
\begin{center}
\resizebox{\textwidth}{!}{\includegraphics{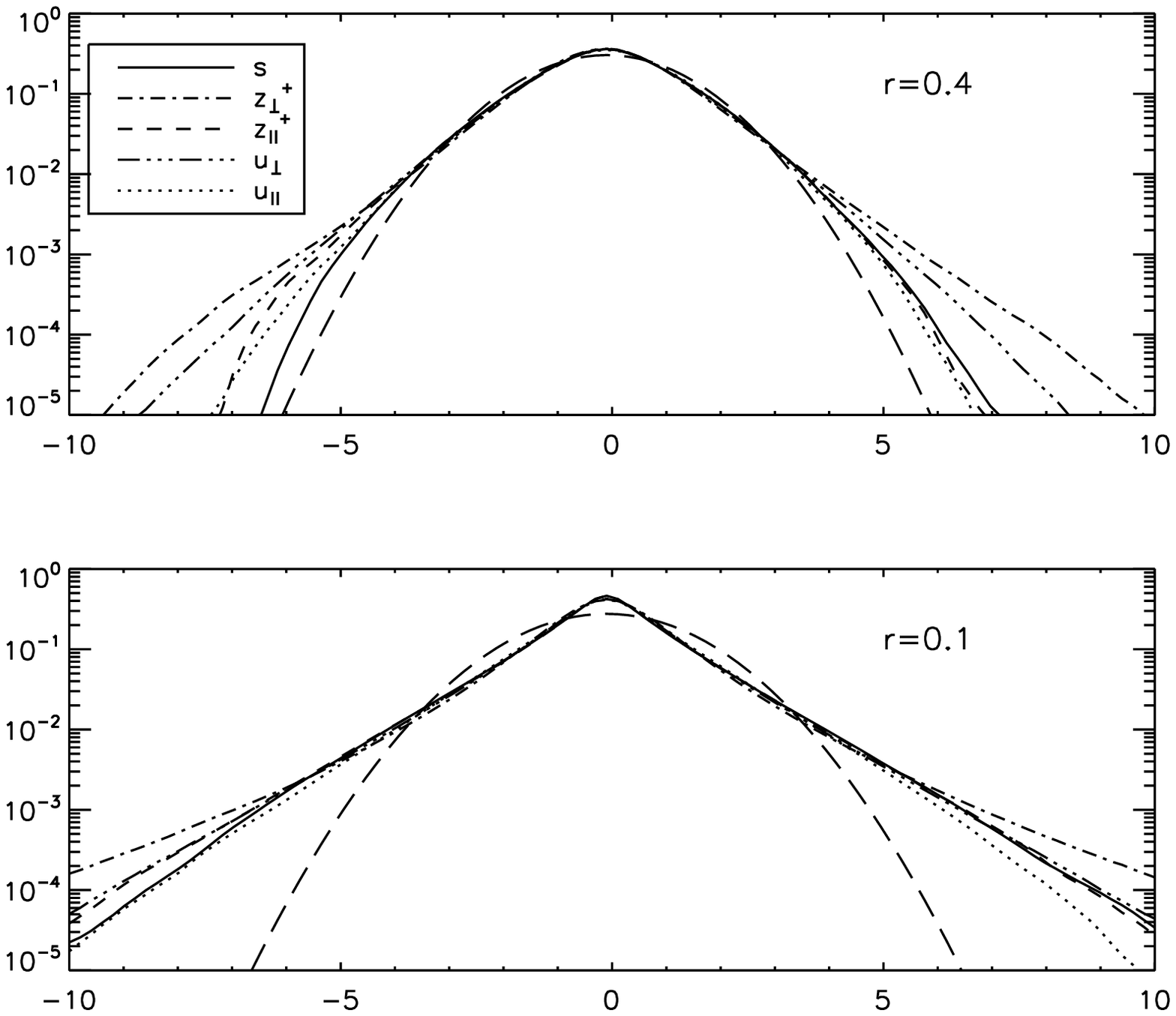}}
\caption{A comparison of the histograms of the normalised scalar differences with those of the shear and pseudo Alfv\'en fields and the field-perpendicular and field-parallel velocity. The long dashed line in both panels is a Gaussian with the same mean and variance as the scalar ($512^3$, $Re=Pe=1800$).}
\label{fig:pdf_scalar_pseudo}
\end{center}
\end{figure}

The histograms of the shear and the pseudo Alfv\'en differences (not shown) are found to exhibit a qualitatively similar $r$-dependent behaviour as for the scalar. However, as shown in figure~\ref{fig:pdf_scalar_pseudo}, the intermittency develops more quickly for the shear Alfv\'en fluctuations, with the scalar and the pseudo Alfv\'en fields being closer to Gaussian distributions at large $r$. In hydrodynamic turbulence the passive scalar is known to be more intermittent than the advecting velocity \cite[see, e.g.,][]{watanabe_g2007}. Within the inertial range we find that kurtosis for the scalar and the field-parallel components of the Els\"asser fields and the velocity are similar (e.g., for $r=0.4$ we find $K=0.8, 1.0, 0.8$ for $\hat{\delta s}$, $\hat{\delta  z^+_\|}$, $\hat{\delta  u_\|}$, respectively). The field-perpendicular velocity is slightly more intermittent ($K=1.5$) and the considerably more intermittent field-perpendicular Els\"asser field ($K=2.2$) is found to be largely due the intermittency of the field-perpendicular magnetic field.

\section{Discussion}
We have investigated the statistical properties of a passive scalar field in driven, incompressible, field-guided MHD turbulence. As well as being a topic of fundamental interest in fluid dynamics, the study was motivated in part by the recent high-resolution numerical simulations and developments in the phenomenological description of MHD turbulence that distinguish the shear Alfv\'en dynamics from the dynamically less significant pseudo Alfv\'en fields and that reveal the presence of a residual energy. In addition, an understanding of passive scalar evolution in magnetised turbulence forms the first step in modelling and interpreting observations of density fluctuations in astrophysics.

A series of direct numerical simulations have led us to draw the following conclusions. First, in the inertial interval the passive scalar obeys an anisotropic version of Yaglom's relation \citep{yaglom_49}. The modified relation (\ref{yaglom2}) was derived analytically and confirmed numerically.  Second, the spectrum of the scalar follows the velocity field spectrum in terms of its anisotropy and wavenumber scaling, although the inertial interval is shorter and a larger statistical ensemble is required in order to exhibit a good scaling. Third, we have quantified the intermittency of the scalar and compared it with the MHD fluctuations. 
Finally, we have demonstrated that the pseudo Alfv\'en mode is in general not a passive scalar, since it is also driven by the pressure gradient term. However, if the large-scale driving of the pseudo Alfv\'en mode is strong enough, it can indeed exhibit a passive scalar spectrum.

\section*{Acknowledgments}
This work was supported in part by the NSF sponsored Center for Magnetic Self-Organization in Laboratory and Astrophysical Plasmas at the University of Chicago and the University of Wisconsin at Madison, by the US DOE award no.~DE-SC0003888, and by the National Science Foundation under grant no.~NSF PHY11-25915. JM and SB appreciate the hospitality and support of the Kavli Institute for Theoretical Physics, University of California, Santa Barbara, where this work was completed. The simulations were made possible through allocations of advanced computing resources provided by the NSF TeraGrid allocation TG-PHY110016 at the National Institute for Computational Sciences, USA, and the UKMHD consortium's system at the University of Warwick.

%\bibliographystyle{aip}
%\bibliography{References_jm}

\end{document}